\documentclass[sigconf]{acmart} %

\usepackage{array}
\newcolumntype{L}[1]{>{\raggedright\let\newline\\\arraybackslash\hspace{0pt}}m{#1}}
\newcolumntype{C}[1]{>{\centering\let\newline\\\arraybackslash\hspace{0pt}}m{#1}}
\newcolumntype{R}[1]{>{\raggedleft\let\newline\\\arraybackslash\hspace{0pt}}m{#1}}

\usepackage{makecell}
\usepackage{tabularx}
\usepackage{tabulary}
\usepackage{multirow}
\usepackage{booktabs}
\newcommand{\tabitem}{~~\llap{\textbullet}~~} %

\AtBeginDocument{%
  \providecommand\BibTeX{{%
    \normalfont B\kern-0.5em{\scshape i\kern-0.25em b}\kern-0.8em\TeX}}}

\copyrightyear{2024} 
\acmYear{2024} 
\setcopyright{rightsretained} 
\acmConference[TechDebt '24]{International Conference on Technical Debt}{April 14--15, 2024}{Lisbon, Portugal}
\acmBooktitle{International Conference on Technical Debt (TechDebt '24), April 14--15, 2024, Lisbon, Portugal}\acmDOI{10.1145/3644384.3644475}
\acmISBN{979-8-4007-0590-8/24/04}

\begin{document}

\title[Different Debt]{Different Debt: An Addition to the Technical Debt Dataset and a Demonstration Using Developer Personality}

\author{Lorenz Graf-Vlachy}
\email{lorenz.graf-vlachy@tu-dortmund.de}
\orcid{0000-0002-0545-6643}
\affiliation{%
  \institution{TU Dortmund University}
  \streetaddress{Otto-Hahn-Straße 4}
  \city{Dortmund}
  \country{Germany}
  \postcode{44227}
}
\affiliation{%
  \institution{University of Stuttgart, Institute of Software Engineering}
  \streetaddress{Universit{\"a}tsstraße 38}
  \city{Stuttgart}
  \country{Germany}
  \postcode{70569}
}
\author{Stefan Wagner}
\email{stefan.wagner@iste.uni-stuttgart.de}
\orcid{0000-0002-5256-8429}
\affiliation{%
  \institution{University of Stuttgart, Institute of Software Engineering}
  \streetaddress{Universit{\"a}tsstraße 38}
  \city{Stuttgart}
  \country{Germany}
  \postcode{70569}
}

\begin{abstract}
\textit{Background:} The ``Technical Debt Dataset'' (TDD) is a comprehensive dataset on technical debt (TD) in the main branches of more than 30 Java projects. However, some TD items produced by SonarQube are not included for many commits, for instance because the commits failed to compile. This has limited previous studies using the dataset. \textit{Aims and Method:} In this paper, we provide an addition to the dataset that includes an analysis of 278,320 commits of all branches in a superset of 37 projects using Teamscale. We then demonstrate the utility of the dataset by exploring the relationship between developer personality by replicating a prior study. \textit{Results:} The new dataset allows us to use a larger sample than prior work could, and we analyze the personality of 111 developers and 5,497 of their commits. The relationships we find between developer personality and the introduction and removal of TD differ from those found in prior work. \textit{Conclusions:} We offer a dataset that may enable future studies into the topic of TD and we provide additional insights on how developer personality relates to TD.
\end{abstract}

\ccsdesc[500]{Software and its engineering~Maintaining software}
\ccsdesc[500]{Software and its engineering~Risk management}
\ccsdesc[500]{Human-centered computing~Open source software}
\ccsdesc[500]{Social and professional topics~Systems development}
\ccsdesc[500]{Social and professional topics~Software maintenance}
\ccsdesc[500]{Social and professional topics~User characteristics}

\ccsdesc[500]{Software and its engineering}
\ccsdesc[500]{General and reference~Empirical studies}

\keywords{technical debt, developer personality, Teamscale}

\maketitle

\section{Introduction}

\subsection{Technical Debt}

The metaphorical notion of technical debt (TD) was introduced by Ward Cunningham more than 30 years ago~\cite{Cunningham.1992}. Despite slight variations in the exact definition of the term, there is a reasonable consensus on TD being a ``collection of design or implementation constructs that are expedient in the short term, but set up a technical context that can make future changes more costly or impossible''~\cite[p. 112]{Avgeriou.2016}. Over time, despite its inherent limitations~\cite{Schmid.2013}, the TD metaphor has gained acceptance and found widespread use in both academic and practitioner circles~\cite{Ramac.2022}.

Researchers have identified different types of TD and made efforts to categorize them. \citeauthor{Tom.2013}, for example, identify eight different dimensions of TD, comprising code, design and architecture, operational processes, among others~\cite{Tom.2013}. Relatedly, \citeauthor{Alves.2016} and \citeauthor{Rios.2018b} each distinguish between 15 different types of TD, including, for example, design debt, code debt, test debt, and documentation debt \cite{Alves.2016, Rios.2018b}.

By definition, TD comes with advantages and disadvantages. Reported advantages lie particularly in increased short-term developer velocity~\cite{Power.2013}. Reported disadvantages~\cite{Besker.2017} are decreased long-term velocity~\cite{Power.2013}, reduced developer morale and motivation~\cite{Ghanbari.2017, Besker.2020, Ramac.2022}, as well as lower code quality and increased uncertainty and risk~\cite{Tom.2013}.

Beyond conceptual considerations~\cite{Schmid.2013}, qualitative studies~\cite{Ghanbari.2017}, and surveys~\cite{Besker.2017}, researchers have increasingly become interested in performing large-scale quantitative studies on TD using data mined from software repositories~\cite{Lenarduzzi.2019c, Lenarduzzi.2019b, Lenarduzzi.2020, Codabux.2020}.

\subsection{The Technical Debt Dataset and its Limitations}

To enable such studies of TD, \citeauthor{Lenarduzzi.2019} developed the ``Technical Debt Dataset'' (TDD)~\cite{Lenarduzzi.2019}. It is a dataset of TD (primarily code debt) in various Apache Software Foundation (ASF) projects written in Java. It has been used in a variety of studies on TD~\cite{Lenarduzzi.2019c, Lenarduzzi.2019b, Lenarduzzi.2020, Codabux.2020}. In its most recent version 2.0, it contains a comprehensive analysis of the main branches of 31 projects.
Aside from data obtained from PyDriller, Refactoring Miner, Jira, and  Ptidej, the dataset most notably includes TD information generated using SonarQube.

Importantly, SonarQube requires the build of a commit to complete before it can provide TD information. For various reasons (e.g., deficient code or missing dependencies), however, builds may fail. Consequently, the TDD does not contain complete information for such commits. According to our own analyses, the SonarQube analyses were incomplete for more than 60\% of commits in the covered projects.

Recent research has found that this is particularly problematic in cases where differences in TD between commits are of interest, because then missing information in either the focal commit or its parent commit leads to missing data in the ultimate analysis. This problem has been reported to reduce the size of samples dramatically, potentially impacting the validity of analyses performed on them. For instance, in the recent study on developer personality and TD by \citeauthor{GrafVlachy.2023}, the authors collected personality data for 121 developers who made commits that are within the scope of the TDD, but they could only use data from 19 developers due to missing TD information in the TDD~\cite{GrafVlachy.2023}.

Further, it is well-known that TD detection tools may come to different assessments of TD~\cite{Lefever.2021}. As the TDD only contains TD items from SonarQube, it is thus naturally limited in this way, too.

\section{An Addition to the Technical Debt Dataset}

To address these limitations, we develop an addition to the TDD that includes information on TD for essentially all commits. The following describes the used tools, the process of constructing the dataset, and the resulting dataset itself.

\subsection{Teamscale}

We develop our addition to the TDD using Teamscale in version 9.1.2. Teamscale is a tool for analyzing code quality and tests~\cite{Haas.2019, Heinemann.2014}. For the Java language, such analyses can be performed directly on the source code without the need for compiled bytecode. Teamscale can be run locally and allows the user to access its analyses through a web interface or a REST API. It has been previously used in research on empirical software engineering~\cite{Niedermayr.2019}. Teamscale is commercial in nature but CQSE, the company developing it, offers free licenses for open-source projects and academic users.

Notably, the philosophy of the company behind Teamscale discourages the use of single-indicator metrics to assess the maintainability of, and thus the TD in, software projects~\cite{Niedermayr.2016}. Consequently, Teamscale does not provide a singular metric of TD comparable to SonarQube's ``technical debt'' metric (variable  \texttt{sqale\_index}) that has been used in prior work~\cite{GrafVlachy.2023}. Instead, Teamscale provides various detailed measures related to TD. These include, for instance, excessive nesting depth, cyclomatic complexity, malformed comments, name shadowing, hard-coded credentials, or unused code.

\subsection{Construction of the Dataset}

We constructed the dataset in the following way. First, we identified the projects included in the TDD. Although version 2 of the TDD only includes 31 projects, we opted to additionally include all projects listed in the original TDD paper (\textit{Accumulo, Ambari, Atlas, Aurora, Beam, MINA SSHD})~\cite{Lenarduzzi.2019}. Similarly, we decided not to restrict our analyses to the projects' main branches as TDD version 2 did but to analyze all branches. We then implemented a Python tool
that performs several steps. First, it clones the repositories locally. It then imports these local copies into Teamscale, which is also running locally. Once a project is successfully imported, Teamscale begins to perform an analysis of all commits in the background. To ensure complete data availability, our tool waits until data processing within Teamscale is completed. The tool then uses Teamscale's REST API to request all available relevant datapoints for each commit in each branch of each project. Finally, the tool writes these datapoints out to the local disk.

We only analyzed Java code. We used Teamscale's default settings except for two cases. First, we ensured that Teamscale would not only analyze the main branch but all branches by enabling ``Branch support''. Second, we switched on the ``Preserve empty commits'' commits option to ensure that Teamscale would retain all commits.

Data analysis took multiple weeks on a dedicated Windows virtual server with eight cores and 192 GB RAM. The analysis script, the Teamscale configuration file, and the resulting dataset are available at \url{https://doi.org/10.6084/m9.figshare.24550840}.

The dataset and all code are licensed under Apache License 2.0.

\subsection{Description of the Dataset}

There are two key elements which constitute our new dataset. For one, there is a folder for each project with JSON files for each commit in the project that includes all information Teamscale has about the respective commit. This is provided only for advanced use cases. The filenames include each commit's hash for easy identification. In the further analyses of this paper, these files will not be used.

For another, there is a set of CSV files that comprise selected TD information on each commit in the projects. Specifically, there are three types of CSV files. First, there is a ``report'' file. This file contains various aggregated pieces of information that Teamscale provides for each commit. This includes, for instance, the number of parent commits, the number of files in the commit, the lines of code in the commit, the number of findings added and removed in the focal commit, and the number of findings above a certain severity level so Teamscale flags them as ``yellow'' or ``red'', respectively. Several of these data points are provided as an absolute value for the focal commit and as a difference to the parent commit. (Note that Teamscale provides difference data even when the parent is on a different branch. Although the behavior in the case of merge commits is not specifically documented, our investigations lead us to believe that Teamscale compares a focal commit's data to its oldest parent commit.) Table~\ref{tab:report} describes the ``report'' file further.

Second, there is a ``findings'' file. It contains the non-aggregated information on all Teamscale findings per commit (as identified by commit hash). This includes 57 different types of findings, categorized into architecture, comprehensibility, correctness, documentation, efficiency, error handling, redundancy, security, structure, testing, and others. The data is separated out by whether the finding was added or removed in the focal commit, or found in changed code, as well as by finding severity (either ``yellow'' or ``red'').

Third, there is a file on ``findings\_messages'', which provides the detailed Teamscale messages for all findings per commit (as identified by commit hash).

All types of files include the project name, the branch name, and the commit hash as identifiers that allow the data to be linked to each other as well as to the TDD. (Note that some projects---e.g., \textit{Accumulo} and \textit{Batik}---have renamed their main branches from ``master'' to ``main'' between the release of the TDD and our analyses.)

The ``report'' output also includes the first commit of the repository. For these commits, Teamscale lists the \texttt{Author Name} as ``Teamscale import'' but reports no further data through the API although the web interface shows analysis reports. However, due to the particular characteristics of these initial orphan commits, it is likely best to discard these commits in analyses anyway. The ``findings'' output does not include any information on these commits.

Each CSV file exists once for each specific project and once in a combined form that covers all projects. Further information and statistics on the dataset are available in a separate document in the data package at \url{https://doi.org/10.6084/m9.figshare.24550840}.

Note that our dataset is more extensive than the TDD in at least three dimensions regarding TD. It covers more projects, more branches, and it spans a timeframe until the end of October 2023.

Insofar as the two datasets overlap, they can be readily linked using the commit hashes.

\begin{table}[ht]
\def\sym#1{\ifmmode^{#1}\else\(^{#1}\)\fi}
\tiny
\caption{Contents of ``report'' CSV file}
\label{tab:report}
    \centering
    \begin{tabularx}{\linewidth}{llX}
        \toprule
        Category & Variable & Notes \\
        \midrule
        \multirow{3}{*}{Identifiers} & \texttt{Project} & Project name \\
         & \texttt{Branch} & Branch name \\
         & \texttt{Commit Hash} & Commit hash \\
        \midrule
        \multirow{4}{*}{Commit data} & \texttt{Num Parent Commits} & Number of parent commits (0 for orphans, 1 for regular commits, >1 for merge commits) \\
         & \texttt{Timestamp} & Teamscale timestamp of commit \\ 
         & \texttt{Author Name} & Name of author \\ 
         & \texttt{Author Email} & Email address of author \\ 
        \midrule
        \multirow{18}{*}{\makecell{Analysis results}} & \texttt{Files\_<val>} & \multirow{18}{*}{\makecell[l]{\tabitem <val> can be ``abs'' or ``diff''\\ \tabitem <color> can be ``g'' (green),\\ ``y'' (yellow), or ``r'' (red) \\ \tabitem ``cnt'' = ``count'' \\ \tabitem ``assm'' = ``assessment''}} \\
         & \texttt{Lines of Code\_<val>} &  \\ 
         & \texttt{Source Lines of Code\_<val>} &  \\ 
         & \texttt{Longest Method Length\_<val>} &  \\ 
         & \texttt{Maximum Nesting Depth\_<val>} &  \\ 
         & \texttt{Change cnt\_<val>} &  \\ 
         & \texttt{Close Coverage\_<val>} &  \\ 
         & \texttt{Line Coverage\_<val>} &  \\ 
         & \texttt{Num Findings Red\_<val>} &  \\ 
         & \texttt{Num Findings Yellow\_<val>} &  \\ 
         & \texttt{Maximum Cyclomatic Complexity\_<val>} &  \\ 

         & \texttt{File Size assm\_<color>\_<val>} &  \\ 
         & \texttt{Method Length assm\_<color>\_<val>} &  \\ 
         & \texttt{Nesting Depth assm\_<color>\_<val>} &  \\ 
         & \texttt{Comment Completeness assm\_<color>\_<val>} &  \\ 
         & \texttt{Cyclomatic Complexity assm\_<color>\_<val>} &  \\ 

         & \texttt{Added Findings cnt} &  \\ 
         & \texttt{Removed Findings cnt} &  \\ 
         & \texttt{Findings in Changed Code cnt} &  \\ 
         \bottomrule
    \end{tabularx}
\end{table}

\section{Developer Personality and Technical Debt Redux}

We demonstrate the utility of our dataset in an exploration of the relationship between developer personality and TD. To do so, we replicate an analysis of developer personality and TD that was hampered by the limitations of the TDD~\cite{GrafVlachy.2023}.

\subsection{Description of Original Study}

In their recent study, ~\citeauthor{GrafVlachy.2023} used the TDD to explore developer personality~\cite{GrafVlachy.2023} in the context of TD. Specifically, they studied the relationship between three broad personality constructs and the introduction and removal of TD. The three personality constructs are the five traits of the Five Factor Model (extraversion, agreeableness, conscientiousness, emotional stability, and openness to experience), the personality characteristic of regulatory focus (comprising promotion focus and prevention focus), and narcissism.
They propose that incurring TD is a form of risk-taking (also see \cite{GrafVlachy.2023b}), and they argue that different personality characteristics relate, through their relationship with risk-taking, to TD.
They find that conscientiousness, emotional stability, openness to experience, and prevention focus are negatively linked to TD. They find no significant results for extraversion, agreeableness, promotion focus, or narcissism. 

To gather developer personality data, \citeauthor{GrafVlachy.2023} surveyed all 1,555 developers having made any commits that are part of the TDD version 2. Importantly, they measured all variables using validated scales
~\cite{Wagner.2020}. 
The five-factor model personality traits were captured using the Ten-Item Personality Measure (TIPI)
~\cite{Gosling.2003}.
Regulatory focus was measured using six items (three for promotion focus and three for prevention focus) from the Regulatory Focus Composite Scale (RF-COMP)
~\cite{Haws.2010}. 
Narcissism was captured using the short version of the Narcissistic Personality Inventory (NPI-16)
~\cite{Ames.2006}. 
Reliability metrics like Cronbach's $\alpha$ were sufficiently high.

\citeauthor{GrafVlachy.2023} also identified developers' age at the time of each commit by capturing developers' age in years and then subtracting the difference between 2022 and the year in which the focal commit was made from the provided age.

After accounting for missing data and implausible values, they obtained complete data on the characteristics of 121 developers.

\subsection{Demonstration Using Our Dataset}

In the following, we describe our analysis using our new dataset. Importantly, we do not theorize ex ante about any individual relationships between personality and TD. Instead, we simply explore the data to see if we find patterns similar to the ones reported by \citeauthor{GrafVlachy.2023}~\cite{GrafVlachy.2023}.

Note that, in contrast to their analysis (which only used the net amount of TD created or removed by a commit), we study the number of TD items (Teamscale ``findings'') that were added in a commit and those that were removed in a commit separately. For comparability, we additionally use the difference between the two values (i.e., the net change) as a third dependent variable.

\subsubsection{Sample}

Our sample is the result of a merge between the TDD and our dataset by commit hash. It is thus restricted to commits made to the main branches of the projects, which also alleviates concerns over the potentially experimental nature of non-main branches. We only consider normal commits and drop merge and orphan commits~\cite{Alfayez.2018} based on information from the TDD. Merge commits do not allow a sensible calculation of changes in TD (due to multiple parent commits) and orphan commits likely have particular characteristics that may distort the analyses. We further obtained the developer personality data collected by \citeauthor{GrafVlachy.2023} and linked it to our newly developed dataset. Overall, our sample comprises 5,497 commits from 111 developers. This is substantially larger than the sample of \citeauthor{GrafVlachy.2023}, who analyzed 2,145 commits from only 19 developers~\cite{GrafVlachy.2023}. Notably, we still cannot analyze all commits because even Teamscale does not provide data for all. This is the case, for instance, for cross-repository commits.

\subsubsection{Analysis Strategy}

We follow the method used by \citeauthor{GrafVlachy.2023}~\cite{GrafVlachy.2023}. This means that we used panel regressions because each developer is observed repeatedly, once for each commit they made. We clustered standard errors at the developer to account for the fact that such multiple commits from the same developer are not statistically independent. In our model, we controlled for developer age at time of commit (from ~\cite{GrafVlachy.2023}) and lines of code (LOC) added and LOC removed (from the TDD as Teamscale does not provide these metrics). To account for unobserved time-invariant aspects of each project (for instance, specific coding conventions), we included dummy variables (fixed effects) for each project.

Notably, for the analyses of the number of added and removed findings, a Poisson estimator would be econometrically appropriate because these dependent variables are counts~\cite{Wooldridge.2010}. However, because this estimator did not converge when analyzing our dataset, we report the results of a random effects panel model instead. Such a model is the appropriate choice for our third dependent variable, the net change in findings. We will focus our interpretation of the results on this dependent variable, also because it allows for a direct comparison with the original study~\cite{GrafVlachy.2023}.

All analyses were performed in Stata 17.0. All analysis scripts are available at \url{https://doi.org/10.6084/m9.figshare.24550840}.

\subsubsection{Findings}

As is evident in Table~\ref{tab:reg}, we find that LOC added and LOC removed are related to the number of added and removed findings in the way one would expect. We further find a positive effect of extraversion on added findings and net change, negative effects of promotion focus and narcissism on removed findings, and a negative effect of age at commit on net change. Surprisingly, only the finding on age at commit is in line with the prior research from \citeauthor{GrafVlachy.2023}~\cite{GrafVlachy.2023}. All findings regarding personality differ. Specifically, we do not reproduce any of \citeauthor{GrafVlachy.2023}'s significant findings, and all our significant findings were not present in their work~\cite{GrafVlachy.2023}.

\begin{table}[ht]
\def\sym#1{\ifmmode^{#1}\else\(^{#1}\)\fi}
\small
\caption{Results of panel regression analyses \label{tab:reg}}
\centering
\begin{tabular}{lccc} 
\toprule

                    &\makecell{Added\\findings}&\makecell{Removed\\findings}&\makecell{Net\\change}\\
\midrule
Extraversion        &        5.85\sym{*}  &        2.97         &        2.88\sym{*}  \\
                    &      (2.27)         &      (1.85)         &      (1.42)         \\
\addlinespace
Agreeableness       &       -4.23         &       -2.37         &       -1.86         \\
                    &      (3.34)         &      (2.12)         &      (3.62)         \\
\addlinespace
Conscientiousness   &       -1.82         &       -0.24         &       -1.58         \\
                    &      (3.43)         &      (3.40)         &      (2.54)         \\
\addlinespace
Emotional stability &       -0.62         &       -2.48         &        1.87         \\
                    &      (2.59)         &      (1.97)         &      (2.51)         \\
\addlinespace
Openness to experience&       -3.32         &       -2.02         &       -1.30         \\
                    &      (3.30)         &      (3.14)         &      (2.48)         \\
\addlinespace
Promotion focus     &        0.63         &       -1.91\sym{*}  &        2.54         \\
                    &      (1.56)         &      (0.87)         &      (1.73)         \\
\addlinespace
Prevention focus    &       -0.46         &       -0.43         &       -0.03         \\
                    &      (0.67)         &      (0.56)         &      (0.56)         \\
\addlinespace
Narcissism          &       -2.12         &       -2.28\sym{*}  &        0.15         \\
                    &      (1.32)         &      (0.94)         &      (1.42)         \\
\addlinespace
Age at commit       &       -0.27         &        0.63         &       -0.89\sym{*}  \\
                    &      (0.50)         &      (0.34)         &      (0.41)         \\
\addlinespace
LOC added           &        0.05\sym{**} &       -0.02\sym{***}&        0.07\sym{***}\\
                    &      (0.02)         &      (0.00)         &      (0.02)         \\
\addlinespace
LOC removed         &       -0.02\sym{*}  &        0.05\sym{**} &       -0.07\sym{***}\\
                    &      (0.01)         &      (0.02)         &      (0.02)         \\
\addlinespace
Constant            &       48.50         &       47.65         &        0.85         \\
                    &     (52.25)         &     (46.49)         &     (33.72)         \\

\midrule
Project fixed effects        &        Yes         &        Yes         &        Yes         \\
Observations        &        5,497         &        5,497         &        5,497         \\
Clusters              &        111             &       111              &       111              \\

\bottomrule

\multicolumn{4}{l}{\footnotesize Dependent variable indicated in top row.}\\
\multicolumn{4}{l}{\footnotesize Table reports coefficients, clustered standard errors in parentheses.}\\
\multicolumn{4}{l}{\footnotesize \sym{*} \textit{p} $<$ 0.05, \sym{**} \textit{p} $<$ 0.01, \sym{***} \textit{p} $<$ 0.001}\\

\end{tabular}
\end{table}

\section{Discussion}

\subsection{Threats to validity}

\subsubsection{Construct validity}

Our measures of TD relies on automated analyses that may not produce perfectly accurate results. Teamscale can be configured extensively, but we use the default settings since we do not have grounds to make a different choice. In particular, to remain consistent across projects, we do not make use of Teamscale's feature to allow for manually identified ``tolerated'' or ``false positive'' findings. Different configurations might lead to different results.
We also use a simple count of findings as our dependent variables, implicitly assuming that every individual finding represents the same amount of TD. Future research might wish to weigh different types of findings differently.
Further, Teamscale largely captures only code debt, but not other types of TD~\cite{Tom.2013, Alves.2016, Rios.2018b}. %

The used personality data may not be perfectly reliable since it is based on self-reports %
using short scales~\cite{Schmidt.1996}.
Finally, developers' personality data was collected after they made the analyzed commits. This time gap might potentially affect the accuracy of the personality data in case personality would change over time~\cite{Calefato.2019}. 

\subsubsection{Internal validity}

Despite following prior work in our selection of control variables, our regressions might suffer from omitted confounding variables, thus limiting the internal validity of our study. Since we use control variables from the TDD, we can also only analyze commits that are from the main branches of the projects. Developers' characteristics may also be related to whether their code is incorporated into the main branch in the first place, which might affect our results.

\subsubsection{External validity}

As a matter of course, our study is restricted to developers of large ASF projects. This limits the generalizability of our results to other contexts, such as smaller or closed-source projects.
Further, although our analyzed sample is much larger than that of prior work~\cite{GrafVlachy.2023}, the overall response rate of developers in the survey capturing personality information is still low, potentially creating sample selection issues.

\subsubsection{Reliability}
Reliability is likely of limited concern. All used personality scales are well-established in psychology.
We provide the script to re-run the Teamscale analyses as well as the dataset. Unfortunately, we cannot share the dataset that includes personality data for obvious privacy reasons.

\subsection{Implications and Conclusion}

First and foremost, our research provides a fine-grained dataset for future studies of TD. Since we also provide the scripts to generate the dataset, future researchers can recreate it with other Teamscale settings however they see fit. In particular, as our dataset fully integrates with the TDD (by linking via commit hash), we enable extensions of prior studies conducted with it.

In terms of practical implications, the findings from our demonstration using the dataset caution practitioners to not overweight results from any single study, such as the original study using the TDD. In fact, we show how an enlarged sample and different measures of TD may yield very different results. In sum, we hope that our empirical findings and dataset spur further research into the link between developer characteristics and TD.

\begin{acks}
We thank Davide Taibi for information on The Technical Debt Dataset and Colin Kolbe for development support. We thank CQSE GmbH for a Teamscale license and for support with setting up the analysis, and Tobias Röhm for helpful hints.
\end{acks}

\bibliographystyle{ACM-Reference-Format}
\bibliography{base}

\end{document}